\documentstyle[preprint,eqsecnum,aps]{revtex}
\begin{document}
\draft
\title{SURVIVAL AND NONESCAPE PROBABILITIES\\
FOR RESONANT AND NONRESONANT DECAY}
\author{G. Garc\'{\i}a-Calder\'on, J.L. Mateos and M. Moshinsky\thanks{%
Member of El Colegio Nacional.}}
\address{Instituto de F\'{\i}sica,\\
Universidad Nacional Aut\'onoma de M\'exico\\
Apartado Postal 20-364, 01000 M\'exico, D.F., M\'exico}
\maketitle

\begin{abstract}
In this paper we study the time evolution of the decay process for a
particle confined initially in a finite region of space, extending our
analysis given recently [Phys. Rev. Lett. {\bf 74} (1995), 337]. For this
purpose, we solve exactly the time-dependent Schr\"odinger equation for a
finite-range potential. We calculate and compare two quantities: i) the
survival probability $S(t)$, i.e., the probability that the particle is in
the initial state after a time $t$; and ii) the nonescape probability $P(t)$%
, i.e., the probability that the particle remains confined inside the
potential region after a time $t$. We analyze in detail the resonant and
nonresonant decay. In the former case, after a very short time, $S(t)$ and $%
P(t)$ decay exponentially, but for very long times they decay as a power law
albeit with different exponents. For the nonresonant case we obtain that
both quantities differ initially. However, independently of the resonant and
nonresonant character of the initial state we always find a transition to
the ground state of the system which indicates a process of `loss of memory'
in the decay.
\end{abstract}

%\preprint{Final version con correccion 13/Feb/96}

%\pacs{03.65.-w,73.40.Gk,23.90.+w}
\newpage
\narrowtext

\section{Introduction \protect\\}

In a recent publication\cite{prl}, we study the time evolution of the decay
process in Quantum Mechanics for a particle initially confined in a finite
region of space. We were able to solve exactly the time-dependent
Schr\"odinger equation for a finite range central potential $V(r)$ by
considering the full resonant spectra of the system using a novel
representation of the time-dependent Green function in terms of resonant
states.

Once we obtain the wave function $\psi(r,t)$ for our problem we calculate
and compare two quantities: i) the survival probability $S(t)$, that is, the
probability that the particle is in the initial state after a time $t$; and
ii) the nonescape probability $P(t)$, defined as the probability that the
particle remains confined inside the potential region after a time $t$. Of
the two notions the survival probability has been more widely used,
presumably because it can be manipulated in a more abstract way, without
referring to an initially confining region of space for the decaying
particle. One finds that the notion of survival probability has been
commonly used in studies of quantum decay\cite{decay} and also in dynamical
studies of quantum chaos\cite{caos}. Actually the survival and nonescape
probabilities correspond to two different notions\cite{2,3}. For instance, a
particle can leave the initial state $\psi(r,0)$ while remaining inside the
potential. In such a case $S(t)$ varies with time while $P(t)$ is always
unity.

In this paper we will analyze in detail resonant and nonresonant decay. The
former refers to the situation where the energy of the initial state is
close to one of the resonances of the system. This case was considered in
Ref. \onlinecite{prl} and here we extend the discussion also to the
nonresonant decay, that is the case where the energy of the initial state is
arbitrary.

A comparison between the survival and nonescape probabilities might be of
particular interest in connection with the problem of the nonexponential
contributions to the time evolution of quantum decay\cite{decay,ggc,nature};
in threshold effects in photoionzation in atomic physics\cite{rac93} and in
radioactive decay of nuclei\cite{nonexp,sag95}; in one dimensional quantum
tunneling\cite{8}; in tunneling dynamics of squeezed states\cite{squeezed},
and in studies of the time evolution of chaotic quantum systems\cite
{caos,chaotic}.

A novel feature of our approach is that the exponential and nonexponential
contributions to the time evolution of decay are presented in terms of the
full set of complex poles and resonant states associated with the outgoing
Green function of the system. The description covers the full time interval
from $0 \leq t < \infty$ and in particular it allows to study the effect of
the far away resonances on the time evolution properties of the system.

We shall discuss the above notions for a symmetrical potential $V(r)$ as
shown in Fig.\ \ref{fig1}, and for simplicity, we shall restrict ourselves
to s-waves. A first question is whether $S(t)$ and $P(t)$ can be obtained in
an analytic and explicit fashion, and the answer is yes, if $\psi (r, 0)$ is
available in an explicit analytic form. This discussion will be carried out
for arbitrary $V(r)$ and $\psi (r,0)$ in section II using an approach by
Garc\'{\i}a-Calder\'on\cite{ggc}. First we provide the stationary Green
function of the problem as an expansion involving the resonant states and
complex poles of the problem and then, with the help of the inverse Laplace
transform, get the time-dependent Green function as an infinite sum of
terms. Each term consists of coefficients involving the overlap of the
initial state and the corresponding resonant state, multiplied by a function
(introduced long ago by Moshinsky\cite{35}) that depends on the complex
energy and the time.

Once the probabilities of survival and nonescape are explicitly determined
comes the interesting question of discussing their behavior and then
comparing $S(t)$ and $P(t)$ for different times. For the resonant case we
obtain that, after a very short time, both quantities decay exponentially
and coincide with each other; however for the nonresonant case in general
differ considerably. Independently of the resonant or nonresonant character
of the initial state we always find a crossover to the ground state of the
system which indicates a process of `loss of memory'. We will also derive
with more detail the main result of Ref. \onlinecite{prl}, namely, that for
long times, $S(t)$ and $P(t)$ behave in a different way: $P(t)\sim t^{-1}$
while $S(t)$ obeys the well known power law decay $\sim t^{-3}$.

To illustrate the approach it is convenient to consider a specific example
and thus in section III we take the case of the delta-function potential, $%
V(r)=b\delta(r-R)$, and a simple analytical expression for the initial state 
$\psi(r,0)=\beta(\kappa) \sin \kappa r$. We shall in particular analyze the
case where the strength of the $\delta$-function potential is very large, 
{\it i.e.} in our units $b>>1$, because then one can derive an approximate
analytical expression for the poles of the Green function. This will allow
us to discuss in section IV what happens to $S(t)$ and $P(t)$ when the
initial state is near resonance and compare it with the situation when the
initial state is off-resonance. Another aspect of interest is the behavior
of $S(t)$ and $P(t)$ at different times and in particular the change from
exponential to a power law behaviour which we shall also illustrate
numerically.

\section{Determination of the survival and nonescape probabilities\protect\\}

The situation that we may confront then is that of a potential $V(r)$, as
drawn in Fig.\ \ref{fig1}, that terminates at $r=R$. Our initial normalized
wave function is restricted to the interval $0\leq r \leq R$. We can then
solve the time-dependent radial Schr\"odinger equation with the potential $%
V(r)$ and the initial condition mentioned above at $t=0$, and denote the
solution by $\psi (r,t)$. With the help of the latter we can define the
survival amplitude $A(t)$\cite{decay,2,3} as 
\begin{equation}
A(t)=\int^R_0 \psi^\ast (r,0)\psi (r,t)dr  \label{1e1}
\end{equation}
\noindent
so that $S(t)=|A(t)|^2$ is the probability of finding at time $t$ the state $%
\psi(r,t)$ at its initial value $\psi(r,0)$, that is, 
\begin{equation}
S(t)=\left |\int^R_0 \psi^\ast (r,0)\psi (r,t)dr \right |^2.  \label{1e2}
\end{equation}
\noindent
On the other hand the probability that the particle does not escape from the
potential is 
\begin{equation}
P(t) =\int^R_0 \psi^\ast(r,t)\psi(r,t)dr,  \label{1e3}
\end{equation}
\noindent
as the wave function $\psi(r,t)$ remains normalized if that was the case for
its original value $\psi(r,0)$. Thus the probability that the particle
escapes is $1-P(t)$. We shall denote $P(t)$ as the nonescape probability.

To derive a procedure for determining $S(t)$ and $P(t)$, we require first to
find a solution of the time-dependent Schr\"odinger equation for a potential
that vanishes beyond a distance, i.e., $V(r)=0; r > R$, (Fig.\ \ref{fig1}),
and the initial condition $\psi(r,0)$. It is convenient first to determine
the time dependent Green function, which we denote by $g(r,r^\prime,t)$ and
satisfies the equation 
\begin{equation}
\bigg[ - {\frac{\partial^2}{\partial r^2}} +V(r)\bigg]\,g(r,r^\prime,t)=i{%
\frac{\partial }{\partial t}} g(r,r^\prime,t),  \label{2e1}
\end{equation}
with the initial condition 
\begin{equation}
g(r,r^\prime,0) =\delta (r-r^\prime),  \label{2e2}
\end{equation}
where we take $\hbar =2M=1$ and hence $E=k^2$.

We proceed then to find the Laplace transform of this Green function which
we denote by $\bar g(r,r^{\prime },s)$ given by 
\begin{equation}
\bar g(r,r^{\prime };s)=\int_0^\infty g(r,r^{\prime },t)e^{-st}dt.
\label{2e4}
\end{equation}
\noindent
Instead of the complex variable $s$ it is convenient to introduce the
variable $k$ obeying, 
\begin{equation}
s=(-ik^2),\,\,\,k\equiv (is)^{1/2},  \label{2e6}
\end{equation}
\noindent
and to define the outgoing Green function $G^{+}(r,r^{\prime };k)$, which is
used normally in the literature\cite{16} through, 
\begin{equation}
G^{+}(r,r^{\prime };k)\equiv i\bar g(r,r^{\prime };s).  \label{2e6a}
\end{equation}
Applying then the transform to both sides of Eq.\ (\ref{2e1}) we obtain,
using Eq.\ (\ref{2e6a}), 
\begin{equation}
\bigg[ {\frac{\partial ^2}{\partial r^2}}+k^2-V(r)\bigg] G^{+}(r,r^{\prime
};k)=\delta (r-r^{\prime }).  \label{2e5}
\end{equation}
Once the Laplace transform (\ref{2e4}) is determined, we can get $%
g(r,r^{\prime },t)$ by the inverse Laplace transform over the Bromwich
contour but, in terms of the variable $k$, the integral is over the
hyperbolic contour $C_0$ of Fig.\ \ref{fig2} and we obtain\cite{ggc,gcp} 
\begin{equation}
g(r,r^{\prime },t)=\frac i{2\pi }\int_{C_0}\,G^{+}(r,r^{\prime };k){\rm e}%
^{-ik^2t}\,\,2kdk.  \label{2e7}
\end{equation}
\noindent
Thus our first objective will be to obtain $G^{+}(r,r^{\prime };k)$.

\subsection{Determination of ${\bf G^+(r,r^\prime;k)}$\protect\\}

Turning our attention back to Eq.\ (\ref{2e1}) we note that for $r\ne
r^{\prime }$, $G^{+}(r,r^{\prime };k)$ satisfies an homogeneous ordinary
differential equation, namely the Schr\"odinger equation of the problem. It
is well known that $G^{+}(r,r^{\prime };k)$ may be written as\cite{16}, 
\begin{equation}
G^{+}(r,r^{\prime };k)=-\frac{\phi (k,r_{<})f(k,r_{>})}{W(k)},  \label{2e8}
\end{equation}
\noindent
where $r_{<}$ indicates the smaller of $r$ and $r^{\prime }$ and $r_{>}$,
the larger. The function $\phi (k,r)$ stands for the regular solution of the
Schr\"odinger equation of the problem with boundary conditions at the
origin, 
\begin{equation}
\phi (k,0)=0,\,\,\,\,\bigg[ \partial \phi (k,r)/\partial r\bigg]_{r=0}=1,
\label{2e9}
\end{equation}
\noindent
and the function $f(k,r)$ is irregular at the origin and is defined by the
condition that for $r>R$ (since the potential vanishes beyond the distance $R
$), it behaves as ${\rm exp}(ikr)$, namely an outgoing wave, which implies
that for at $r=R$ it satisfies, 
\begin{equation}
f(k,R)=e^{ikR},\,\,\,\,\bigg[\partial f(k,r)/\partial r\bigg]%
_{r=R}=ike^{ikR},  \label{2e10}
\end{equation}
which also defines it completely. The function $W(k)$ represents the
Wronskian, 
\begin{equation}
W(k)=\phi (k,r)\frac{\partial f(k,r)}{\partial r}-f(k,r)\frac{\partial \phi
(k,r)}{\partial r}.  \label{2e13}
\end{equation}
\noindent

It is well known\cite{16} that for a finite range interaction the outgoing
Green function $G^+(r,r^{\prime};k)$, as a function of $k$, can be extended
analytically to the whole complex $k$-plane where it has an infinite number
of poles, that arise from the zeros of the Wronskian (\ref{2e13}), and are
distributed in a well known manner\cite{16,18}. Purely imaginary poles on
the upper half $k$-plane correspond to the bound states of the problem,
whereas purely imaginary poles seated on the lower half $k$-plane correspond
to antibound states. On the other hand complex poles are only found on the
lower half $k$-plane. To each complex pole at $k_n=a_n-ib_n$, with $a_n, b_n
>0$, there corresponds, due to time reversal invariance, a complex pole $%
k_{-n}$ situated symmetrically with respect to the imaginary axis, i.e. $%
k_{-n}=-k_n^*$. In the examples that we shall discuss later, we shall assume
potentials with no attractive part so they have no bound states, and thus
all the poles of $G^+(r,r^\prime;k)$ will be in the lower half of the
complex $k$-plane. Hence our analysis is restricted to resonant states.

As shown below the complex poles of the outgoing Green function are related
to the resonant states of the system. Resonant states may be defined as
solutions of the Schr\"odinger equation of the problem\cite
{gamow,siegert,gcp} 
\begin{equation}
{\frac{d^2u_n(r)}{dr^2}}+\left [ k_n^2-V(r)\right ] u_n(r)=0,  \label{r1}
\end{equation}
\noindent
that obey the usual boundary condition at the origin, 
\begin{equation}
u_n(0)=0,  \label{r2}
\end{equation}
\noindent
and at distances, $r \geq R$, describe a situation where there are no
incident particles, namely, $u_n(r) \propto {\rm exp}(ik_nr)$, and hence
satisfy the outgoing boundary condition, 
\begin{equation}
{\left [\frac{du_n(r) }{dr }\right]}_{r=R_-}=ik_nu_n(R).  \label{r3}
\end{equation}
\noindent
Gamow\cite{gamow} showed that the outgoing boundary condition (\ref{r3})
implies that the energy eigenvalues of the problem are necesarily complex,
namely, $k_n^2=E_n=\epsilon_n-i\Gamma_n/2$, where $\epsilon_n$ stands for
the position of the resonance and $\Gamma_n$ refers to the corresponding
width. The above boundary conditions also provide the bound and antibound
solutions corresponding, respectively, to imaginary positive, $k_n=i\gamma_m$%
, and imaginary negative, $k_n=-i\delta_n$, values of $k_n$, where both $%
\gamma_n$ and $\delta_n$ are real. The above complex eigenvalues correspond
precisely to the complex poles of the outgoing Green function of the problem 
$G^+(r,r^{\prime};k)$. Moreover, it turns out that resonant states may also
be defined as the residues at the complex poles $k_n$ of $G^+(r,r^{\prime};k)
$\cite{res}. Here we follow Garc\'{\i}a-Calder\'on\cite{ggc} to write, 
\begin{equation}
R(r,r^\prime,k_n) = \lim_{k\to k_n}\bigg\{ (k-k_n) G^+(r,r^\prime,k)\bigg\}={%
\frac{ u_n(r) u_n(r^{\prime}) }{2k_n}}  \label{2e20}
\end{equation}
provided the resonant states normalized according to 
\begin{equation}
\int_0^Ru_n^2(r)dr + {\frac{i }{2k_n}}u_n^2(R) =1.  \label{r4}
\end{equation}

Let us now consider the integral\cite{ggc} 
\begin{equation}
{\frac{1}{2\pi i}} \int_{C^{\prime}} {\frac{G^+(r,r^\prime;z)}{z-k}} dz=0,
\label{2e17}
\end{equation}
which is taken over the contour $C^{\prime}$ of Fig.\ \ref{fig3} consisting
of one large cycle, whose radius will eventually go to $\infty$, and small
circles surrounding all the poles of $G^+(r,r^\prime;z)$ in the $z$ plane as
well as the point where $z=k$. As the integrand is analytic inside this
contour the integral vanishes, as indicated on the right hand side of (\ref
{2e17}).

For the large circle, when $|z|\to \infty$, we can, in the upper half $I_+$
of the $z$ plane, essentially disregard the potential $V(r)$ in Eq.\ (\ref
{2e5}), and thus $G^+(r,r^\prime;z)$, obtained from (\ref{2e8}), when we
replace $k$ by $z$, becomes 
\begin{mathletters}
\label{2e18ab}
\begin{equation}
z^{-1}\bigg\{ e^{iz(r+r^\prime)}- e^{iz (r^\prime-r)} \,\bigg\}, \,\, \,\,{%
\hbox{\rm if}}\;\; 0\leq r\leq r^\prime \leq R,  \label{2e18a}
\end{equation}
\begin{equation}
z^{-1}\bigg\{ e^{iz(r+r^\prime)} - e^{iz (r-r^\prime)}\, \bigg\}, \,\, \,\, {%
\hbox{\rm if}}\;\; 0\leq r^\prime\leq r\leq R,  \label{2e18b}
\end{equation}
which clearly vanishes, at least as $|z|^{-1}$ when $z=x+iy$, with $y>0$.
For the lower half $I_-$ of the $z$ plane, and the real axis, the analysis
is more complicated. Garc\'{\i}a-Calder\'on and Berrondo\cite{gcb} have
shown though, using appropriate forms of the Born approximation, that $%
G^+(r,r^\prime;z)$ vanishes there exponentially provided $(r, r^{\prime}) < R
$. Thus there is no contribution from the large circle in Fig.\ \ref{fig3},
and considering the contributions from all the small circles, including the
one around $z=k$, one obtains\cite{ggc,more,romo,bang,gc}, 
\end{mathletters}
\begin{equation}
G^+(r,r^\prime;k)=\sum^\infty_{n=-\infty}{\frac{u_n(r)u_n(r^{\prime}) }{%
2k_n(k-k_n)}};\,\,\, (r,r^{\prime}) < R.  \label{2e19}
\end{equation}
\noindent
Substitution of Eq.\ (\ref{2e19}) into Eq.\ (\ref{2e5}) yields after
straightforward algebra the relations\cite{ggc,gc} 
\begin{equation}
\sum_{n=-\infty}^{\infty}{\frac{u_n(r)u_n(r^{\prime}) }{k_n}}=0\,; \hskip%
.5truecm (r,r^{\prime})<R,  \label{r5}
\end{equation}
\noindent
and 
\begin{equation}
{\frac{1 }{2}}\sum_{n=-\infty}^{\infty} u_n(r)u_n(r^{\prime})=\delta
(r-r^{\prime}); \hskip.5truecm (r,r^{\prime})<R.  \label{r6}
\end{equation}
\noindent
Eqs.\ (\ref{r6}) have been also derived following an expansion of $%
G^+(r,r^{\prime};k)$ in terms of inverse powers of $k$\cite{more,romo}.
Notice that 
\begin{equation}
{\frac{1 }{2k_n(k-k_n)}} \equiv {\frac{1 }{2k}}\left [ {\frac{1 }{k-k_n}} + {%
\frac{1 }{k_n}} \right ],  \label{r7}
\end{equation}
\noindent
so then using Eq.\ (\ref{r5}) one may write Eq.\ (\ref{2e19}) as 
\begin{equation}
G^+(r,r^\prime,k)=\frac {1}{2k}\sum^\infty_{n=-\infty}{\frac{%
u_n(r)u_n(r^{\prime}) }{(k-k_n)}};\,\,\,(r,r^{\prime}) < R.  \label{r8}
\end{equation}
\noindent

\subsection{Determination of ${\bf g(r,r^\prime,t)}$\protect\\}

One way to determine $g(r,r^{\prime},t)$ is deforming the contour $C_0$ of
Fig.\ \ref{fig2} to exploit the analytical properties of $G^+(r,r^{\prime};k)
$ using the theorem of residues. In this form one may obtain expansions
involving a sum of exponentially decaying terms plus an integral
contribution along certain path of the complex $k$-plane\cite{gcp,sud}. If
the potential vanishes beyond a distance and the notions of interest are
defined within the potential region, as happens for the survival and
nonescape probabilities defined by Eqs.\ (\ref{1e2}) and $(\ref{1e3})$, we
can proceed as follows. First we deform the contour $C_0$ to the contour $C$
as shown in Fig.\ \ref{fig2}. This can be easily done because ${\rm exp}%
(-ik^2t)$ converges as $k$ increases in the upper left quadrant of the $k$%
-plane. Since we have assumed absence of bound states we then may write Eq.\
(\ref{2e7}) as, 
\begin{equation}
g(r,r^\prime,t)=\frac{i}{2\pi} \int_{-\infty}^{\infty} \, G^+(r,r^\prime;k) 
{\rm e}^{-i k^2t}\,\,2kdk.  \label{2e7p}
\end{equation}
\noindent
We can then substitute Eq.\ (\ref{r8}) into Eq.\ (\ref{2e7p}) to obtain an
expansion of $g(r,r^{\prime};t)$ in terms of the resonant states of the
problem. The only integral we need to determine is 
\begin{equation}
{\frac{i}{2\pi}} \int^\infty_{-\infty} {\frac{{\rm e}^{-ik^2t}}{k-k_n}} dk,
\label{2e31}
\end{equation}
\noindent
Moshinsky\cite{25} has discussed the above type of integral. It appears in
the description of transient effects in time dependent problems in Quantum
Mechanics\cite{kleber94,Holland} and it may be denoted as, 
\begin{equation}
M(k_n,t) \equiv {\frac{i}{2\pi}} \int^\infty_{-\infty} {\frac{{\rm e}%
^{-ik^2t}}{k-k_n}}dk = \frac{1}{2} {\rm e}^{u^2} {\hbox {\rm erfc}}(u),
\label{2e32}
\end{equation}
\noindent
where 
\begin{equation}
u= -{\rm exp}(-i\pi/4)k_n t^{1/2}.  \label{2e33}
\end{equation}
\noindent
Thus the time dependent Green function $g(r,r^\prime,t)$ takes the form
first derived by Garc\'{\i}a-Calder\'on\cite{ggc}, 
\begin{equation}
g(r,r^{\prime},t)=\sum_{n=-\infty}^{\infty} u_n(r)u_n(r^{\prime})M(k_n,t)\,, %
\hskip.5truecm (r,r^{\prime})<R  \label{2e34}
\end{equation}

Having obtained our basic result, we proceed to the determination of $%
\psi(r,t)$ for an arbitrary initial condition $\psi(r,0)$.

\subsection{Expansion of ${\bf \psi(r,t)}$\protect\\}

If our initial state is $\psi(r,0)$ instead of $\delta (r-r^\prime)$, it is
clear that at time $t$ $\psi(r,t)$ is given by the integral 
\begin{equation}
\psi(r,t) = \int^R_0 \psi(r^\prime, 0) g(r,r^\prime, t) dr^\prime.
\label{2e36}
\end{equation}
We define now $C_n$ and ${\bar C}_n$ by the expressions 
\begin{equation}
C_n \equiv \int^R_0 \psi (r,0) u_n(r) dr;\,\,\, {\bar C}_n=\int_0^R
\psi^*(r,0)u_n(r)\,dr,  \label{2e37}
\end{equation}
\noindent
where, respectively, $u_n(r)$ is normalized according to Eq. (\ref{r4}) and $%
\psi(r,0)$ as, 
\begin{equation}
\int^R_0 |\psi(r,0)|^2 dr=1.  \label{2e37p}
\end{equation}
\noindent
Finally we can write $\psi(r,t)$ as 
\begin{equation}
\psi(r,t) = \sum^\infty_{n=-\infty}C_n u_n(r) M(k_n,t);\,\,\,\,(r < R).
\label{2e39}
\end{equation}
The coefficients $(\ref{2e37})$ obey some useful relations. Multiply both
Eqs.\ (\ref{r5}) and $(\ref{r6})$, respectively by $\psi(r,0)$ and $%
\psi^*(r^{\prime},0)$ and integrate the result from the origin up to the
radius $R$, to obtain\cite{ggc}, 
\begin{equation}
\sum_{n=-\infty}^{\infty}{\frac{C_n {\bar C}_n }{k_n}}=0,  \label{19}
\end{equation}
\noindent
and 
\begin{equation}
{\frac{1 }{2}}\sum_{n=-\infty}^{\infty} C_n {\bar C}_n=1.  \label{20}
\end{equation}

\subsection{Expansion of the survival and nonescape probabilities \protect\\}

From (\ref{1e1}), (\ref{2e37}) and (\ref{2e39}) we get that 
\begin{equation}
A(t) = \sum^\infty_{n=-\infty} C_n{\bar C}_n M (k_n,t),  \label{2e40}
\end{equation}
\noindent
It is worth mentioning that as long ago as 1951, Moshinsky had given an
exact expression of $A(t)$ for the one-level case, i.e., when only $k_{\pm
1} \neq 1$, see Eq. (27b) of ref. \cite{25}. Substitution of Eq.\ (\ref{2e40}%
) into Eq.\ (\ref{1e2}) leads to the following expression for the survival
probability, 
\begin{equation}
S(t)= \sum_{n=-\infty}^{\infty} \sum_{\ell=-\infty}^{\infty} C_n {\bar C}_n
C^{\ast}_{\ell} {\bar C}^{\ast}_{\ell} M (k_n,t) M^{\ast}(k_{\ell},t).
\label{2e41}
\end{equation}
\noindent
Using $\psi(r,t)$, as given by (\ref{2e39}), one may obtain an expression
for the nonescape probability $P(t)$ with the help of (\ref{1e3}). We only
need to add the definition 
\begin{equation}
I_{n \ell} = \int^R_0 u^*_{\ell}(r) u_n(r) dr  \label{2e42}
\end{equation}
to obtain 
\begin{equation}
P(t)= \sum_{n=-\infty}^{\infty} \sum_{\ell=-\infty}^{\infty} C_n
C^{\ast}_{\ell}I_{n \ell} M (k_n,t) M^{\ast}(k_{\ell},t).  \label{2e43}
\end{equation}

Thus, we have the general expressions for the probabilities of survival and
nonescape in the form of a double sum involving the products $M(k_n,t)
M^\ast (k_\ell, t)$ of functions defined by (\ref{2e32}), with coefficients
of the type $C_n, {\bar C}_n, I_{n\ell}$ given respectively by (\ref{2e37})
and (\ref{2e42}). Notice that the only difference between both expansions is
that in $P(t)$, $I_{n\ell}$ replaces the product ${\bar C}_n^*{\bar C}%
_{\ell}^*$ that appears in $S(t)$.

\subsection{Exponential and long time behaviour\protect\\}

Eqs,\ (\ref{2e40}), $(\ref{2e41})$ and $(\ref{2e43})$ are given in terms of $%
M$ functions and therefore their exponential behaviour is not exhibited
explicitly. This can be achieved by using the symmetry relations between the
poles on the third and fourth quadrants, $k_{-p}=-k_p^*$, to write the sums
only over the poles $k_p$ located on the fourth quadrant and make use of the
relation \cite{ggc,15}, 
\begin{equation}
M(k_p,t)={\rm e}^{-ik_p^2t}-M(-k_p,t),  \label{12}
\end{equation}
\noindent
where $M(-k_p,t)$ is defined as in Eq.\ (\ref{2e32}) but with the argument $%
u= {\rm exp}(-i\pi/4)k_n t^{1/2}$ instead of $(\ref{2e33})$. Using the above
relation the survival amplitude $(\ref{2e40})$ may be written as 
\begin{equation}
A(t) = \sum^\infty_{p=1} C_p{\bar C}_p {\rm e}^{-ik_p^2t}-I(t),  \label{s1}
\end{equation}
\noindent
that displays explicitly the exponentially decaying behaviour recalling that 
$k_p^2=\epsilon_p-i\Gamma_p/2$. The term $I(t)$ on the right-hand side of
the above equation stands for the nonexponential contribution and it reads, 
\begin{equation}
I(t)= \sum^\infty_{p=1}\left [ C_p{\bar C}_p M(-k_p,t)] - C_p^*{\bar C}_p^*
M(-k^*_p,t) \right ],  \label{s2}
\end{equation}
\noindent
where we have used that $C_{-p}{\bar C}_{-p}$=$C_p^*{\bar C}_p^*$ and $%
k_{-p}=-k_p^*$. The argument of the function $M(-k_p^*,t)$ is $u={\rm exp}%
(-i\pi/4)k_p^*t^{1/2}$. Notice that the sums in Eqs.\ (\ref{s1}) and $(\ref
{s2})$ run over the same set of poles. The nonexponential contribution given
by Eq.\ (\ref{s2}) appears as an infinite sum of terms on the same footing
that the description of the exponentially decaying terms. Is is worth
mentioning that Khalfin, who was the first author to discuss nonexponential
contributions to decay, referred to models involving a single resonance term%
\cite{khalfin}. In other treatments, the nonexponential contribution appears
as a complicated integral term, and in general has been treated in an
approximate way\cite{decay,sud,peres,26}

If one ignores the nonexponential contribution $I(t)$ in Eq.\ (\ref{s1}) and
furthermore assumes that the initial state is very close to the sharpest
resonant state, say the $sth$, then from (\ref{19}) and (\ref{20}) it
follows that $C_s\bar C_s \approx 1$, all other coefficients being very
small, then (\ref{s1}) may be written as $A(t) = {\rm e}^{-ik_s^2t}$ and the
survival probability becomes the well known expression, 
\begin{equation}
S(t) = {\rm e}^{-\Gamma_s t}.  \label{s3}
\end{equation}
\noindent
A similar analysis can be done for the nonescape probability. Noticing that $%
I_{ss} \approx 1$, as follows from inspection of (\ref{r4}) and (\ref{2e42}%
), one obtains 
\begin{equation}
P(t) = {\rm e}^{-\Gamma_s t}.  \label{s4}
\end{equation}
\noindent
The equivalence of $S(t)$ and $P(t)$ as shown above, along the exponentially
decaying region, has probably led to confusion regarding the notions of
survival and nonescape probabilities.

Let us now turn to the analysis of the long time behavior, as the discussion
for very short times has been given elsewhere\cite{15}. From Eq.\ (\ref{12})
one sees that asymptotically the relevant terms are of the type $M(-k_n,t)$
where $k_n$ stands for either $k_p$ or $k_p^*$. Indeed from the definition
of the function $M$, given by (\ref{2e32}), one may get the asymptotic
expansion of ${\rm exp}(u^2){\rm erfc}(u)$ to obtain\cite{ggc,15} 
\begin{equation}
M(-k_n,t) \approx \frac{i}{2(\pi i)^{1/2}} \left(\frac {1}{k_nt^{1/2}}\right
)- \frac{1}{4(\pi i)^{1/2}} \left(\frac {1}{k_n^3t^{3/2}}\right )+... .
\label{22}
\end{equation}
\noindent
Substitution of Eq.\ (\ref{22}) into $(\ref{s2})$ allows to write $I(t)$ as
a sum over terms that go like inverse powers of time. One may see, however,
that the coefficient proportional to $t^{-1/2}$, is identical to Eq.\ (\ref
{19}) and therefore it cancels out exactly. Consequently the survival
amplitude $(\ref{s1})$ may be written as, 
\begin{equation}
A(t) \approx \sum_{p=1}^{\infty} \left [ C_p \bar{C}_p{\rm e}^{-ik_p^2t}- 
\frac{1}{4(\pi i)^{1/2}} \left [{\frac{ C_p\bar{C}_p }{k_p^3}} - \left ({%
\frac{ C_p\bar{C}_p }{k_p^3}} \, \right )^* \right ] {\frac{1 }{t^{3/2}}}%
+..\right ].  \label{23}
\end{equation}
\noindent
The above equation exhibits the crossover from exponential to a power law
behaviour. Using $(\ref{23})$ one sees that the survival probability $(\ref
{2e41})$ behaves at long times as $S(t) \sim t^{-3}$.

Let us now consider the long time behaviour of the nonescape probability.
One sees from Eq.\ (\ref{2e34}) into Eq.\ (\ref{1e3}) for $P(t)$, that the
resonant expansion of $g(r,r^{\prime},t)$ is coupled through the integration
over $r$, with that of $g^*(r,r^{\prime},t)$. This originated the integrals $%
I_{n \ell}$ defined by $(\ref{2e42})$. Hence, when the asymptotic expansion
of the $M$ functions $(\ref{22})$ is introduced into $(\ref{2e43})$, the
leading contribution is proportional to $t^{-1}$ and includes terms of the
type 
\begin{equation}
\sum_r^{\infty}\sum_s^{\infty} \left( \frac {C_r^*C_s I_{rs}}{k_r^*k_s}
\right ) {\frac{1 }{t}},  \label{24}
\end{equation}
\noindent
that are different from $(\ref{19})$ and hence do not cancel. In other
words, asymptotically, $P(t) \sim t^{-1}$. One concludes therefore, that the
survival and the nonescape probabilities behave with a different inverse
power of time for long times\cite{prl}.

The previous discussion refers only to resonant states, however it can be
easily extended to include bound states\cite{ggc}. In such a case the long
time behaviour of the survival and nonescape probabilities has a
nonvanishing asymptotic value due to the oscillating character of the time
dependence of the bound states. For instance, if in addition to resonant
states there is a bound state, as time goes to infinity the survival
probability reads, 
\begin{equation}
S(t)=|C_b|^2,  \label{25}
\end{equation}
where $C_b$ and $\bar C_b=C_b^*$ stand for the overlap intregrals defined in
Eq.\ (\ref{2e37}) with the $u_n(r)$ replaced by the bound state function $%
u_b(r)$. The above nonvanishing asymptotic value for the survival
probability is sometimes referred to in the literature as population trapping%
\cite{rac93}. Proceeding in a similar fashion as above for the nonescape
probability yields, in the limit that the time goes to infinity, that 
\begin{equation}
P(t)=|C_b|^2.  \label{26}
\end{equation}
One sees that both $S(t)$ and $P(t)$ provide at very long times a similar
result when bound states enter into the problem.

\section{Example\protect\\}

We shall now consider a problem that illustrates numerically the behaviour
of $S(t)$ and $P(t)$. We take $V(r)$ as a delta-function potential 
\begin{equation}
V(r) = b\delta (r-1),  \label{1e4}
\end{equation}
\noindent
with strength $b$ and where we choose $R=1$. For the initial condition we
take 
\begin{equation}
\psi(r,0) = \cases{\beta (\kappa) \sin \kappa r, &if $0 \leq r \leq 1,$\cr 0
& if $r > 1,$}  \label{1e5}
\end{equation}
where $\kappa$ is an arbitrary real parameter and, for normalization, we
require that 
\begin{equation}
\beta (\kappa) = \sqrt{2} [1-(2\kappa)^{-1} \sin 2\kappa]^{-1}.  \label{1e6}
\end{equation}
The form (\ref{1e5}) was suggested by the fact that if $b \rightarrow \infty$%
, the normalized stationary states for the problem are 
\begin{equation}
\psi(r,0) = \cases{\sqrt{2} \sin m \pi r, &if $0 \leq r \leq 1 ;
m=1,2,3,\dots$\cr 0 & if $r > 1,$}  \label{3e1}
\end{equation}
and so an interesting question could be what happens at any time $t$ if at $%
t=0$ we consider the above initial state. This state implies that initially
we are very close to a resonance, and since we would like to consider also
the case of an initial state situated between two resonances, we use the
more general form (\ref{1e5}).

From (\ref{r1}) we immediately see that 
\begin{equation}
u_n(r) = d_n sin k_nr,  \label{3e2}
\end{equation}
\noindent
where $d_n$ and the equation satisfied by $k_n$ will be given below. Using (%
\ref{3e2}) we see that Eq.\ (\ref{r3}) has to be modified to take into
account the fact that the delta-function potential is discontinuous at $r=R=1
$. Hence for this example instead of Eq.\ (\ref{r3}) we have the condition 
\begin{equation}
ik_nu_n(1)-{\left[\frac{ du_n(r) }{dr }\right]}_{r=R_-=1_-}=bu_n(1).
\label{3e3p}
\end{equation}
\noindent
Substitution of Eq.\ (\ref{3e2}) into the above expression yields the
equation for the complex eigenvalues of the problem, namely, 
\begin{equation}
2ik_n + b({\rm e}^{2ik_n}-1)=0.  \label{3e4}
\end{equation}
The solutions to the above trascendental equation correspond also to the
poles $k_{n}$, $(n=\pm 1, \pm 2, \pm 3, \dots)$ of $G^+(r,r^{\prime};k)$ and
in this case all of them are in the lower half of the $k$ plane with $%
k_{-n}=-k^\ast_n$.

The coefficient $d_n$ in Eq. \ (\ref{3e2}) may be obtained from the
normalization condition given by Eq.\ $(\ref{r4})$. Using (\ref{3e4}) it may
be written as, 
\begin{equation}
d_n=\left [{\frac{2(b-2ik_n) }{(1+b-2ik_n)}} \right ]^{1/2}.  \label{3e3}
\end{equation}

In order to calculate the survival and nonescape probabilities, given
respectively by (\ref{2e41}) and (\ref{2e43}), we need to determine also the
coefficients $C_n$ and $I_{n\ell}$ using (\ref{3e2}) for $u_n(r)$ and (\ref
{1e5}) for $\psi(r,0)$. Since the initial condition $\psi
(r,0)=\beta(\kappa) \sin \kappa r$ depends on the parameter $\kappa$, we
shall denote $C_n$ as $C_n(\kappa)$. Hence from (\ref{2e37}) we obtain 
\begin{equation}
C_n (\kappa)= \beta(\kappa)d_n\,{\frac{k_n{\rm e}^{-ik_n} }{b(\kappa^2-k_n^2)%
}}\, \left [ \kappa \cos \kappa-i(k_n+ib)\sin \kappa \right ].  \label{3e5}
\end{equation}
\noindent
Note that as our initial state (\ref{3e2}) is real $C_n=\bar C_n$. On the
other hand from (\ref{2e42}) and (\ref{3e2}) 
\begin{equation}
I_{n\ell}=i{\frac{d_nd_{\ell}^*k_nk_{\ell}^* }{b^2(k_{\ell}^*-k_n) }} {\rm e}%
^{i (k_{\ell}^*-k_n)},  \label{3e7}
\end{equation}
where for all of these coefficients we have made use of the fact that, from (%
\ref{3e4}), $k_n$ satisfies 
\begin{equation}
e^{2ik_n} = \frac{2k_n+ib}{ib}.  \label{3e8}
\end{equation}

Substitution of the coefficients given by (\ref{3e3}), $(\ref{3e5})$, and $(%
\ref{3e7})$ in (\ref{2e41}) and (\ref{2e43}), leads to explicit analytic
expressions for the probabilities of survival $S(t)$ and nonescape $P(t)$
that we now proceed to discuss.

An interesting case refers to sharp isolated resonances. For not very highly
excited resonances, this case can be achieved by taking a large value for
the intensity of the potential, namely, in our units, $b \gg 1$. In that
case, as shown in ref. \cite{15}, the poles of $G^+(r,r^{\prime};k)$ are
given approximately by 
\begin{equation}
k_n \simeq n\pi \bigg( 1-{\frac{1}{b}}\bigg) - i \bigg( {\frac{n\pi}{b}}%
\bigg)^2.  \label{4e7}
\end{equation}

\subsection{ Comparison of ${\bf S(t)}$ and ${\bf P(t)}$\protect\\}

We shall analyze the survival and nonescape probabilities when $b\gg 1$ and $%
\kappa = m\pi$. The complex energy $E_n$ associated with $k_n$ of (\ref{4e7}%
) is approximately given by 
\begin{equation}
E_n = k^2_n \simeq n^2\pi^2 - 2 i {\frac{n^3\pi^3}{b^2}},  \label{4e9}
\end{equation}
so that the separation of the resonances, {\it i.e.} the real part of the
difference $E_{n+1}-E_n$, is much larger than the width of one of them, {\it %
i.e.} the imaginary part of $E_n$. The choice $\kappa=m\pi$ implies that the
initial state (\ref{1e5}) is associated with an energy $\kappa^2=m^2\pi^2$
and thus it is very close in energy to the corresponding resonance as
follows from (\ref{4e9}) with $n=m$. Since $\beta(m\pi)=(2)^{1/2}$, we see
that the formula (\ref{3e5}) for $C_n(m\pi)$ becomes 
\begin{equation}
C_n(m\pi) \approx \frac {2nm}{b(m^2-n^2)};\,\,\, n \neq m,  \label{4e10}
\end{equation}
\noindent
that has a large factor $b$ in the denominator. On the other hand for $n=m$,
we have from ({\ref{4e7}), that $(m^2\pi^2 -k_m^2) \approx (2m^2\pi^2/b)$
and thus 
\begin{equation}
C_m(m\pi) \approx 1.  \label{4e11}
\end{equation}
}

We can also analyze for $b \gg 1$, and $b$ much larger than $n$ and $\ell$,
the behaviour of $I_{n\ell}$. We get from (\ref{3e7}) and (\ref{4e7}) that 
\begin{equation}
I_{n\ell} \approx \frac {2i n\ell\pi}{b^2(\ell-n)};\,\,\, n \neq \ell,
\label{4e12}
\end{equation}
\noindent
and 
\begin{equation}
I_{nn} \approx 1  \label{4e13}
\end{equation}
Thus, if $b \gg 1$, and also $b$ much larger than $m$, and the initial state
is very close to the $mth$ resonant state, we see from (\ref{4e10}), (\ref
{4e11}) and (\ref{4e13}) that the survival and nonescape probabilities
essentially reduce to 
\begin{equation}
S(t) \approx P(t) \approx {\rm e}^{-\Gamma_m t}= {\rm e}^{-t/\tau},
\label{4e14}
\end{equation}
\noindent
with $\tau$ defined as 
\begin{equation}
\tau = \frac {b^2}{4m^3\pi^3}.  \label{4e15}
\end{equation}
\noindent
Hence when we are close to a narrow resonance, the survival and nonescape
probabilities coincide for a number of lifetimes until the contribution of
the resonance with the largest lifetime dominates the time evolution of $S(t)
$ and $P(t)$, as will be shown in the numerical examples presented in the
next section.

On the other hand when the initial state is between resonances, i.e. $\kappa
=(m+1/2)\pi$, inspection of Eq.\ (\ref{3e5}) yields 
\begin{equation}
C_n((m+1/2)\pi) \approx -\frac {2in}{(m^2-n^2)\pi};\,\,\, n \neq m,
\label{4e16}
\end{equation}
\noindent
and for $n=m$ 
\begin{equation}
C_n((n+1/2)\pi) \approx -\frac {2i}{\pi},  \label{4e17}
\end{equation}
\noindent
which indicates that there is no a single coefficient $C_n$ that dominates
over the expansions for $S(t)$ or $P(t)$. Hence there is no reason to expect
initially an exponential decay law behaviour. In the next section we also
exemplify this situation. \noindent

\subsection{ Numerical results \protect\\}

In order to analyze $S(t)$ and $P(t)$ we need to specify two parameters: the
strength $b$ of the delta-function potential and the wavenumber $k$ for the
initial condition. In what follows, we will take the fixed value $b=100$. On
the other hand, we have to find the poles $k_n$ in the complex $k$-plane,
that satisfy Eq.\ (\ref{3e4}). In order to do so, we use the well-known
Newton-Raphson method to locate the poles in the fourth quadrant $k_n$ $%
(n=1,2, \dots, N)$; the remaining poles lie on the third quadrant and are
given by $k_{-n}=-k^\ast_n$. We have considered in our calculations 1000
poles (N=500); those on the fourth quadrant are depicted in Fig.\ \ref{fig4}%
. Notice that the real part of $k_n$ is much larger than the imaginary part
of $k_n$ and, as a consequence, we are in a regime in which the real part of
the complex energy $E_n$ is much larger than its imaginary part, that is, we
are dealing with sharp isolated resonances. The poles in Fig.\ \ref{fig4}
depend on the parameter $b$, and when $b>>1$ they are given approximately by
Eq.\ (\ref{4e7}). In fact, we use (\ref{4e7}) as an input for the
Newton-Raphson algorithm, in order to obtain the exact location of the
poles. When $b\to\infty$, the poles ``move'' to the real axis in the $k-$%
plane and are given by $k_n=n\pi$; this limiting case corresponds to the
bound states of a particle in a box.

Once we have obtained the analytical expression for the survival $S(t)$ and
nonescape $P(t)$ probabilities (Eqs.\ (\ref{2e41}) and (\ref{2e43})), we can
analyze both quantities numerically. As mentioned above we will take $b=100$%
. In the remaining figures we plot the logarithm of $S(t)$ as a solid line
and the logarithm of $P(t)$ as a dashed line. In Fig. 5 we show ${\rm ln}%
\,S(t)$ and ${\rm ln}\,P(t)$ as a function of time, when the initial
condition has a value $\kappa =5\pi $. This value corresponds to an initial
state close to the fifth resonance and hence refers to the time evolution of
an initial state at a higher excitation energy to those considered in Ref.
[1], where the initial states were considered, respectively, near the first
and second resonances of the system. The fifth resonance has a lifetime $%
\tau _5=1/\Gamma _5$, where $\Gamma _5$ is its width. We will take the time
in Figs. 5 and 6 in units of $\tau _5$.

We see from Fig. 5a that at the beginning of the decay process, both
quantities coincide and both decay exponentially with a lifetime $\tau_5$,
as expected from the choice of the initial condition. As is well known\cite
{15} the exponential decay is not valid for very small times, i.e., times
much smaller than a lifetime, however this deviation cannot be appreciated
in our Figures due to the scale. In Fig. 5b we can see that after a time $%
\approx 10 \tau_5$, $S(t)$ and $P(t)$ separate from each other; $S(t)$
decays exponentially with a slope $\Gamma_5$ for about 20 lifetimes ($%
t\approx 20 \tau_5)$. After that time, something interesting happens: $S(t)$
starts to oscillate and then decay exponentially once again, but this time
with a lifetime $\tau_1$ which corresponds to the ground state of the
system. The nonescape probability $P(t)$ , after a time $\approx 10 \tau_5$,
change the slope of the exponential decay from $\Gamma_5$ to $\Gamma_1$, but
in a smooth way without any oscillations. Finally, in Fig. 5c, we show the
behavior of $S(t)$ and $P(t)$ from the beginning of the decay process until
very long times $(t/\tau_5=6000)$ in order to analyze the asymptotic decay.
We can clearly see three separate regions for both $S(t)$ and $P(t)$ : the
exponential region with slope $\Gamma_5$ at the beginning; the crossover to
an exponential decay with slope $\Gamma_1$ corresponding to the ground state
and; another crossover (for $\tau/\tau_5 \approx 4000)$ to a power law decay
in which $S(t) \sim t^{-3}$ and $P(t)\sim t^{-1}$. Notice that $P(t)$ is
always a smooth function and that $S(t)$, on the other hand, fluctuates
during the transitions from one regime to another.

Before we analyze the nonresonant case, let us try to understand the results
depicted in Fig. 5. First of all, when we choose an initial condition $%
\psi(r,0)$ such that $k=m\pi$, with $m$ an integer, we are very close to a
resonance given by the pole $k_m$. In this case we have a leading term for $%
S(t)$ and $P(t)$ due to the resonant denominator of the form $%
(m^2\pi^2-k_m^2)^{-1}$. The presence of this leading term guarantee an
exponential decay with a lifetime $\tau_m =\Gamma^{-1}_m$, associated with
the complex energy $E_m=\epsilon_m-i\Gamma_m/2$. Furthermore, as discussed
above, when $k=m\pi$ and $b>>1, P(t)$ and $S(t)$ coincide. This is precisely
what we found in the example shown in Fig. 5a, when $m=5$.

Now, in order to understand the crossover to the ground state with a
lifetime $\tau_1=\Gamma_1^{-1}$, we have to remember that in our example
there exist an ordering $\tau_1> \tau_2> \dots >\tau_N$, that is, the ground
state is the state with the longest lifetime and therefore it dominates the
decay process at long times. In terms of the widths, the first resonance is
the thinnest. The fluctuations obtained for $S(t)$ in the interval 20$\tau_5$
to 200$\tau_5$ in Fig. 5b, correspond to the interplay of all the states
between $m=1$ and $m=5$. In addition the the cases $m=1$ and $m=2$
considered in Ref. \onlinecite{prl}, we have also analyzed the case in which 
$m=3$, (not shown here) where $S(t)$ decays exponentially with a lifetime $%
\tau_3$ at the beginning and, during the crossover to the ground state, it
oscillates periodically with a period given by $(\epsilon_2 -
\epsilon_1)^{-1}$.

Finally, as discussed for the general case of a finite range potential
(section II), the asymptotic behaviors of $S(t)$ and $P(t)$ are governed by
the asymptotic form of the function $M(k_n,t)$ and the sum rules given by (%
\ref{19}) and (\ref{20}). The result is that we obtain numerically for the
delta-function potential that $S(t) \sim t^{-3}$ and $P(t)\sim t^{-1}$, as
expected.

Let us now turn to the nonresonant case in which $\kappa $ is not a multiple
of $\pi $. We choose $\kappa =5.5\pi $, which corresponds to an initial
state situated between two resonances, i.e., $m=5$ and $m=6$. In Fig. 6 we
show ${\rm ln}\,S(t)$ and ${\rm ln}\,P(t)$ as a function of time, in units
of $\tau _5$. Once again, the dashed line correspond to ${\rm ln}\,P(t)$ and
the solid line to ${\rm ln}\,S(t)$. In Fig. 6a we can see that at the
beginning of the decay process neither $S(t)$ nor $P(t)$ decay
exponentially. $P(t)$ is a smooth curve but $S(t)$ fluctuates much more
strongly than the resonant case during the entire process. For longer times $%
(t/\tau _5\approx 100)$, both quantities start to decay exponentially with a
lifetime $\tau _1$, as can be seen in Fig. 6b, even though we are in the
off-resonance case characterized by $\kappa =5.5\pi $. Finally, in Fig. 6c
we can see a crossover, for very long times $(t/\tau _5\approx 3000)$, to a
power law decay. Once again, in this regime $S(t)\sim t^{-3}$ and $P(t)\sim
t^{-1}$.

Let us try to understand the differences between $P(t)$ and $S(t)$ from a
physical standpoint. The nonescape probability $P(t)$ is, by definition, the
probability that a particle remains confined inside a certain region. Thus,
the particle can change from the initial state to another state inside the
potential and this change does not affect $P(t)$. The only way in which $P(t)
$ can decrease is when the particle leaves the potential region, say by
tunneling. In the problem we are analyzing, once outside, the particle
cannot enter this region again and, therefore, we expect a decreasing smooth
curve describing the nonescape probability. On the other hand, the survival
probability $S(t)$ refers to the probability that a particle remains after a
time $t$ in its initial state, where the latter is confined inside the
potential region. Suppose that a particle is initially between two
resonances. Before the particle leaves the potential region, it can make
transitions from one state to another and return to the initial state an so
on. All these transitions are manifested by the rapid oscillations of the
survival probability.

A result of our analysis that seems of interest to stress here is that
independently on the conditions for the initial state eventually the
resonant state with the longest lifetime predominates over the time
evolution of decay. Apparently the system looses memory on whether the
initial state is close to the $mth$ or the $nth$ resonance, or between the $%
mth$ and the $nth$ resonances. The last stage of the decaying process
proceeds always exponentially before the croosover to the power law decay
takes over. Indeed, at the beginning of the decay proccess the differences
in the time evolution for different initial states may be dramatic, as shown
for example by Figs. 5a and 6a, and also by Figs. 5b and 6b. However once
the transition to the first resonance occurs the decay out of this state and
the crossover to the power law behaviour becomes similar as shown by Figs.
5c and 6c.

The above situation may be understood in general by inspection of Eq.\ (\ref
{20}) in view of (\ref{2e41}) and (\ref{2e43}) for $S(t)$ and $P(t)$,
respectively, and using the relation (\ref{12}) to exhibit the exponential
dependence of the above definitions explicitly. The initial state determines
the values of the different coefficients $C_n\bar C_n$ and the main point is
to realize that for the longest lived state, say the $sth$, even if the
coefficient $C_s\bar C_s$ is small, the product $C_s\bar C_s {\rm exp}%
(-ik_s^2t)$ will eventually dominate over all other exponentially decaying
terms. This follows because the other decaying terms, having smaller
lifetimes, go to zero much faster.

\section{Concluding remarks\protect\\}

In this work we have made use of an exact expansion for the retarded time
dependent Green function in terms of resonant states to study the time
evolution of an initially confined arbitrary state. Our approach allows to
treat on the same footing both the exponential and nonexponential
contributions to the time evolution of decay. We have compared the notions
of the survival and nonescape probabilities and found that in general they
exhibit a different behaviour with time. For the case of an initial state
close to a resonance the above notions coincide along the exponentially
decaying region. However, if the initial state is not close to a resonance
these probabilities exhibit a nonexponential behaviour. In the above two
cases, we have shown that eventually the resonance with the longest lifetime
predominates in the decay process which may be seen as a `loss of memory' of
the initial condition. This occurs in an exponential fashion before the
onset to a power law decay is established, i.e. $S(t) \sim t^{-3}$ and $P(t)
\sim t^{-1}$. However in the case that bound states are present both
quantities tend to a constant, i.e., $S(t)\sim P(t) \sim |C_b|^2$.

The above formulation may be generalized to bidimensional systems, in
particular to systems with arbitrary geometries\cite{caos}. Also, our
formulation may be of interest in studies of the tunneling dynamics of
squeezed states\cite{squeezed} and in the time evolution of chaotic quantum
systems, where nonexponential decay have also been obtained \cite
{33,weiden91,weiden95}. Here it could be of interest, for example, to
investigate the role of the power law decay for the case of quantum systems
that behave chaotically in the classical limit. Furthermore, our results
might be of interest in the efforts to verify experimentally the
nonexponential contributions to decay\cite{nature}. For nuclear radioactive
decay it has been reported\cite{nonexp} no deviation from exponential decay
up to 45 half-lives for the case of $Mn^{56}$. A more promising possibility
to observe nonexponential contributions to decay could be in photodetachment
near threshold in atomic physics\cite{rac93,haan92,pir90}. The analysis of
the present work could be of interest in that field because negative ions in
photodetachment near threshold are usually modelled by a single electron in
a short range potential.

This work was partially supported by DGAPA-UNAM No. IN-101693. We thank K.W.
McVoy for useful discussions.

\begin{figure}[tbp]
\caption{Short range potential $V(r)$ between $r=0$ and $r=R$. The dashed
areas indicate schematically two resonant states, which are centered around
an energy $E_i$ and have a width $\Gamma_i$.}
\label{fig1}
\end{figure}

\begin{figure}[tbp]
\caption{The Bromwich contour $C_o$ in the $k-$plane for the integral given
by eq. (2.10), which can be modified to the real axis $C$, due to the factor 
$\exp(-ik^2t)$, which forces the integral to vanish on the portion of the
circle indicated by the dashed line. The complex poles of the Green function
(2.11) found on the lower half $k-$plane are also indicated.}
\label{fig2}
\end{figure}

\begin{figure}[tbp]
\caption{The contour $C^\prime$ in the $z-$plane for the integral given by
eq. (2.17), consisting of one large circle and small circles surrounding all
the poles of $G^+(r,r^\prime,z)$ as well as $z=k$.}
\label{fig3}
\end{figure}

\begin{figure}[tbp]
\caption{Location of the first 500 poles $k_n$, given by eq. (3.7), in the
complex $k-$plane. Only the fourth quadrant is shown and notice that ${\rm Re%
}(k) \gg {\rm Im}(k)$.}
\label{fig4}
\end{figure}

\begin{figure}[tbp]
\caption{Logarithm of the survival (solid line) and non-escape (dashed line)
probabilities as a function of time, for $b=100$ and the resonance case $%
k=5\pi$. The time scale is given in units of $\tau_5$. (a) At the beginning
of the decay process both quantities coincide and decay exponentially with a
lifetime $\tau_5$. The slope is given by $\Gamma_5=1/\tau_5$. (b) After a
time $\simeq 10 \tau_5$, $S(t)$ and $P(t)$ separate from each other, and
both probabilities start to decay with a lifetime $\tau_1$, which
corresponds to the ground state. Notice the oscillations for $S(t)$ during
the crossover from the excited state (slope $\Gamma_5$) to the ground state
(slope $\Gamma_1$). (c) For very long times (6000 $\tau_5$) we can clearly
see three separate regions for both $S(t)$ and $P(t)$: the exponential
region with slope $\Gamma_5$ at the beginning; the crossover to the ground
state with slope $\Gamma_1$; and finally, another crossover to a power law
decay, in which $S(t)\sim t^{-3}$ and $P(t)\sim t^{-1}$.}
\label{fig5}
\end{figure}

\begin{figure}[tbp]
\caption{Logarithm of the survival (solid line) and non-escape (dashed line)
probabilities as a function of time, for $b=100$ and the nonresonance case $%
k=5.5\pi$. The time scale is given in units of $\tau_5$. (a) In contrast to
the resonance case (see Fig. 5a), neither $S(t)$ nor $P(t)$ decay
exponentially, and $S(t)$ displays a fluctuating character. (b) Although we
are in the off-resonance case, after some time, $S(t)$ and $P(t)$ start to
decay with a lifetime $\tau_1$, which corresponds to the ground state. (c)
For very long times ($\approx$ 4000 $\tau_5$) we can distinguish three
separate regions for $S(t)$ and $P(t)$: a non exponential region at the
begining; the crossover for the ground state with slope $\Gamma_1$; and
finally, the transition to a power law decay, in which $S(t) \sim t^{-3}$
and $P(t)\sim t^{-1}$.}
\label{fig6}
\end{figure}

\end{document}